# Determination of crystal structure and physical properties of $Ru_2Al_5$ intermetallic from first-principles calculations

Jing Luo[a], Meiguang Zhang[b,*] , Xiaofei Jia[a], Xuanmin Zhu[c], Qun Wei[a,*]

[a]*School of Physics, Xidian University, Xi'an 710071, China*

[b]*College of Physics and Optoelectronic Technology, Baoji University of Arts and Sciences, 721016 Baoji, China*

[c]*School of Information, Guizhou University of Finance and Economics, Guiyang 550025, China*

**ABSTRACT**

Novel ordered intermetallic compounds have stimulated much interest. Ru–Al alloys are a prominent class of high-temperature structural materials, but the experimentally reported crystal structure of the intermetallic $Ru_2Al_5$ phase remains elusive and debatable. To resolve this controversy, we extensively explored the crystal structures of $Ru_2Al_5$ using first-principles calculations combined with crystal structure prediction technique. Among the calculated X-ray diffraction patterns and lattice parameters of five candidate $Ru_2Al_5$ structures, those of the orthorhombic *Pmmn* structure best aligned with recent experimental results. The structural stabilities of the five $Ru_2Al_5$ structures were confirmed through formation energy, elastic constants, and phonon spectrum calculations. We also comprehensively analyzed the mechanical and electronic properties of the five candidates. This work can guide the exploration of novel ordered intermetallic compounds in Ru–Al alloys.

**Keywords:** Crystal structure, Mechanical properties, First-principles calculations

[*] Corresponding authors.
 *E-mail addresses:* zhmgbj@126.com (M. Zhang), qunwei@xidian.edu.cn (Q. wei)

# Introduction

Ru–Al alloys have attracted have attracted considerable interest as advanced high-temperature materials with superior thermodynamic properties, strong oxidation and corrosion resistance, high melting points, and high mechanical strength [1-4]. Lu and Pollock [5] studied the ductility and deformation behaviors of Ru-Al alloys in five independent slip systems at room temperature. Since the pioneering studies of Obrowski, who characterized intermetallic Ru-Al using optical microscopy, X-ray diffraction (XRD) analysis, and thermal studies [6,7], the phase diagram of Ru-Al alloys has been experimentally modified and updated [8-10]. Prins *et al.* [11] conducted a theoretical study of Ru-Al alloys, obtaining results consistent with the experimental findings. Five intermetallic compounds have been consensually identified in the Ru-Al binary system: $RuAl$, $Ru_2Al_3$, $RuAl_2$, $Ru_4Al_{13}$, and $RuAl_6$. Using multiple measurement methods and techniques, Mi *et al.* [12] re-examined the Al-rich region of the Ru-Al phase diagram, revealing an orthorhombic $Ru_2Al_5$ intermetallic phase formed at 1492 °C via a peritectic reaction of $RuAl_2$ with the liquid. However, the atomic positions of this new $Ru_2Al_5$ phase could not be precisely determined from the inexact XRD profiles, so only the lattice parameters were provided. Accordingly, the $Ru_2Al_5$ intermetallic phase was assumed to be isostructural to $Fe_2Al_5$ (*Cmcm*, Z= 2). Using an alternative mechanical alloying and subsequent heat treatment, Bai *et al.* [13] prepared the mother alloy $Al_{79}$:$Ru_{21}$ of a skeletal Ru catalyst. They reported that the $Ru_2Al_5$ phase appeared in all samples after heat treatment at 550 °C and disappeared after 2 h of heat treatment at 700 °C. Despite these efforts, the atomic sites of $Ru_2Al_5$ remain uncharacterized. Motived by previous experimental works [12,13], Wen *et al.* [14] analyzed six Ru-Al intermetallic compounds using first-principles calculations. However, the calculated lattice constants of the supposed $Fe_2Al_5$-type $Ru_2Al_5$ intermetallic phase deviated from recent experimental results [12]. A Subsequent theoretical study proposed a hexagonal $P6_3/mmc$-$Ru_2Al_5$ [15]. As clarifying structure of $Ru_2Al_5$ is essential for properly understanding Ru-Al alloys, we conducted a systematic crystal structure search of $Ru_2Al_5$ using the particle swarm optimization (PSO) algorithm in CALYPSO code [16-18]. Among several candidate $Ru_2Al_5$ structures identified by the search algorithm, the orthorhombic *Pmmn* structure most favorably agreed with the available experimental data. The structural stabilities and mechanical and thermodynamic properties of the obtained $Ru_2Al_5$ structures were systematically analyzed.

# Computational methods

The PSO algorithm is the global optimization technique used in CALYPSO. By updating the positions of particles based on both individual and collective optimal solutions, the PSO algorithm effectively explores complex search spaces to predict stable and metastable crystal structures [16]. The CALYPSO code has successfully predicted the crystal structures of various systems [19-23]. In the present study, the energetically favorable structures of $Ru_2Al_5$, were identified through variable-cell crystal structure simulations of $Ru_2Al_5$ at 0 GPa. During the structure search, the population size was set to 50 in each generation, and the first generation was randomly generated under symmetry constraints. Thirty generations were set to achieve convergence. Once the structure search had terminated, the energetically stable structures were retained for further structural relaxations and related property simulations in the Vienna *Ab* initio Simulation Package [24]. The generalized gradient approximation was adopted for the projector-augmented wave ion–electron potentials [25] and the Perdew–Burke–Ernzerhof exchange-correlation functional [26]. A 400 eV cutoff was set for the plane wave expansion and the grid spacing of the Monkhorst–Pack *k* meshes [27] was $2\pi \times 0.04$ $Å^{-1}$ to ensure good convergence of the total energy ($1 \times 10^{-5}$ eV/atom). The single-crystal elastic constants were deduced from the strain-stress relations [28]. The polycrystalline elastic moduli were calculated by the Voigt-Reuss-Hill averaging method [29]. The dynamic stabilities were evaluated from the phonon spectra obtained in PHONOPY code, employing the finite displacement method [30].

# Results and discussion

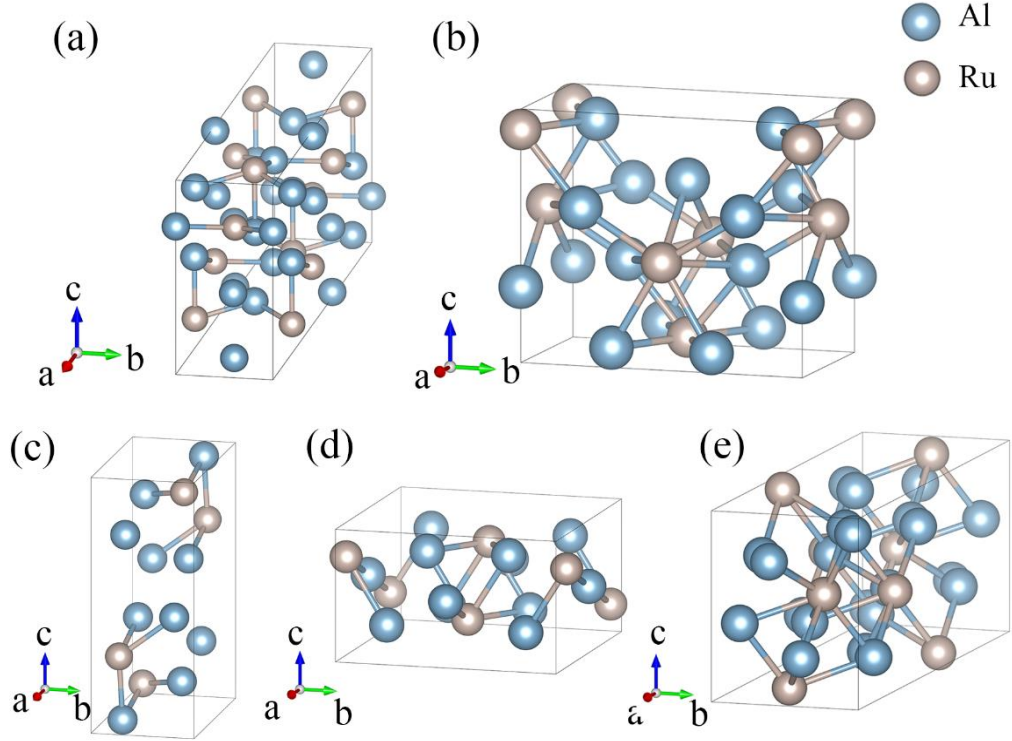

**Fig. 1.** Crystal structures of the (a) *C*2/*m*-1, (b) *Pmmn*, (c) *P*2$_1$/*m*, (d) *C*2/*m*-2, and (e) *C*2/*c* phases of Ru$_2$Al$_5$.

The crystal structures were comprehensively searched at ambient pressure using CALYPSO code. The search obtained five energetically stable Ru$_2$Al$_5$ structures, one orthorhombic and four monoclinic phases (Fig. 1). In the *C*2/*m*-1 phase, the Al and Ru atoms are linearly arranged along the [010] and [110] directions. Along the [001] direction and two arrangements appear along the [001] direction: an alternating sequence of two Al atoms and one Ru atom and a linear sequence of Al atoms. The *P*2$_1$/*m* phase presents two types of zigzag chains along the [100] direction: a chain of alternating Al and Ru atoms and a chain comprised only Al atoms. Along the [100] and [010] directions of the *Pmmn* phase, both the Al and Ru atoms are arranged in straight lines. In the *C*2/*m*-2 phase, the Al and Ru atoms are arranged in straight lines along the three principal axes. Table 1 lists the lattice constants and atomic positions of the Ru$_2$Al$_5$ structures at ambient pressure after full structural relaxations. The lattice constants of the discovered *Pmmn* phase ($a$ = 4.235 Å, $b$ = 7.612 Å, and $c$ = 6.637 Å) are remarkably consistent with the experimental data provided by Mi *et al.* ($a$ = 4.2 Å, $b$ = 7.8 Å, and $c$ = 6.6 Å) [12]. The thermodynamic stability of each Ru$_2$Al$_5$ intermetallic phase was determined from the formation energy ($\Delta H$) [31] of each phase as follows:

$$\Delta H = [E(\text{Ru}_2\text{Al}_5) - 2E(\text{Ru}) - 5E(\text{Al})]/7, \tag{1}$$

Herein, $E(Ru_2Al_5)$ denotes the total free energy of each structure at 0 K, and $E(Ru)$ and $E(Al)$ represent the energies of Ru and Al atoms in metals, respectively. Here, we also considered the previously proposed *Cmcm* $Fe_2Al_5$-type and hexagonal $P6_3/mmc$ phases [12,15] of $Ru_2Al_5$. The $\Delta H$ values of all structures are negative (Fig. 2), suggesting that they are thermodynamically stable and potentially synthesizable. The *Pmmn* phase has the second lowest energy among the considered structures and a lower energy than the earlier reported $Fe_2Al_5$-type and $P6_3/mmc$ phases. Therefore, the discovered *Pmmn* structure is probably the actual structure of $Ru_2Al_5$ reported in previous experimental works [12,13]. Fig. 3 plots the phonon spectra of the five $Ru_2Al_5$ structures. The eigenfrequencies of the lattice vibrations in the Brillouin zone of each structure are positive at ambient pressure, indicating the dynamic stability.

**Table 1**

Space group (SG), lattice parameters, and atomic Wyckoff positions of the five new $Ru_2Al_5$ phases.

| SG | Lattice parameters | | | | Atomic Wyckoff positions | |
|---|---|---|---|---|---|---|
| | $a$ (Å) | $b$ (Å) | $c$ (Å) | $\beta$ (°) | Al | Ru |
| *C2/m*-1 | 13.119 | 4.126 | 7.990 | 106.1 | 4*i* (0.999, 0.0, 0.761) | 4*i* (0.805, 0.0, 0.150) |
| | | | | | 4*i* (0.589, 0, 0.631, 0.631) | 4*i* (0.891, 0.5, 0.693) |
| | | | | | 4*i* (0.899, 0.5, 0.028) | |
| | | | | | 4*i* (0.689, 0.5, 0.528) | |
| | | | | | 4*i* (0.696, 0.5, 0.177) | |
| *Pmmn* | 4.235 | 7.613 | 6.637 | | 4*e* (0.5, 0.713, 0.376) | 2*a* (0.0, 0.0, 0.919) |
| | | | | | 4*e* (0.0, 0.322, 0.078) | 2*b* (0.0, 0.5, 0.424) |
| | | | | | 2*a* (0.0, 0.0, 0.299) | |
| *P2$_1$/m* | 5.086 | 4.138 | 9.968 | 98.9 | 2*e* (0.855, 0.750, 0.446) | 2*e* (0.620, 0.750, 0.898) |
| | | | | | 2*e* (0.510, 0.750, 0.621) | 2*e* (0.055, 0.750, 0.703) |
| | | | | | 2*e* (0.673, 0.750, 0.167) | |
| | | | | | 2*e* (0.141, 0.750, 0.964) | |
| | | | | | 2*e* (0.797, 0.250, 0.749) | |
| *C2/m*-2 | 6.724 | 7.536 | 4.380 | 92.9 | 8*j* (0.661, 0.685, 0.756) | 4*i* (0.163, 0.0, 0.259) |
| | | | | | 2*d* (0.5, 0.0, 0.5) | |
| *C2/c* | 14.506 | 4.699 | 5.910 | 90.5 | 8*f* (0.718, 0.379, 0.819) | 8*f* (0.856, 0.283, 0.529) |
| | | | | | 8*f* (0.923, 0.885, 1.384) | |
| | | | | | 4*e* (0.0, 0.607, 0.750) | |

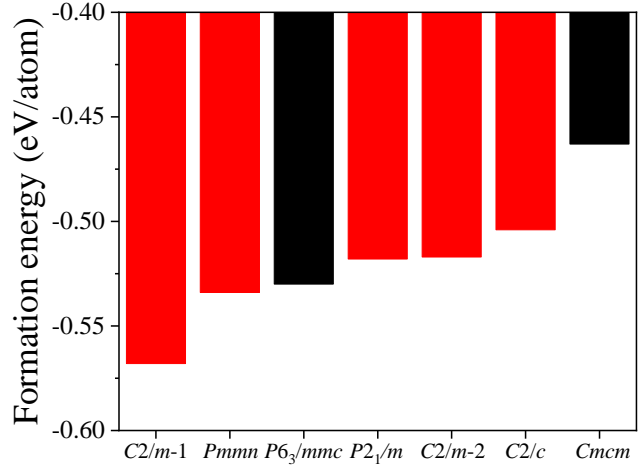

**Fig. 2.** Calculated formation energies of different $Ru_2Al_5$ phases. Red bars represent the five new phases.

Inspired by the perfect consistency between the predicted and experimental lattice constants of the *Pmmn* phase and the energy advantage of *Pmmn* phase over the previously proposed $Fe_2Al_5$-type and $P6_3/mmc$ phases, we further confirmed the *Pmmn* phase of $Ru_2Al_5$. To this end, we compared the XRD patterns of this *Pmmn* phase and that obtained in recent experiments. The simulated and experimental XRD patterns of the *Pmmn* structure are presented in Fig. 4. Our simulations adopted Cu-*Kα* radiation for consistency with earlier experiments. The obtained main peaks of the *Pmmn* structure almost coincided with the experimental data, implying that the experimental phase is indeed the *Pmmn* phase [12]. It should be noted that a lower-energy stable structure (*C*2/*m*-1 phase; Fig. 2) was also identified. Therefore, if the $Ru_2Al_5$ structure of the *Pmmn* phase is the $Ru_2Al_5$ structure synthesized by Mi *et al.* [12], then *Pmmn* phase is likely a metastable phase.

The elastic parameters are fundamental for understanding the mechanical performances of Ru-Al intermetallic phases. The elastic constants $C_{ij}$ of the five $Ru_2Al_5$ structures are listed in Table 2. A necessary condition of crystal stability is mechanical stability. For monoclinic systems (*C*2/*m*-1, *P*2$_1$/*m*, *C*2/*m*-2, and *C*2/*c*) and the orthorhombic *Pmmn* structure, mechanical stability can be assessed under mechanical stability criteria [32-34]. Based on our calculations, all five structures met the criteria, confirming their mechanical stability. The elastic constants of *C*2/*m*-1 and *Pmmn* trend as: $C_{11} < C_{22} \approx C_{33}$ and $C_{11} > C_{22} = C_{33}$, respectively, indicating lower anisotropy of axial incompressibility in these structures than in the other candidate strucutres. $C_{44}$ and $C_{66}$ refer to the shear-resistance deformations under a shear stress along the <001> direction of the (100) plane and under a shear load along the <110> direction of the (100) plane, respectively. As $C_{44} > C_{66}$ in both the *Pmmn* and *C*2/*c* phases, the shear modulus of deformation is larger along the <001> direction than along the <110> direction of (100) plane. In the remaining three structures, $C_{44} < C_{66}$, implying opposite shear behaviors on the (100) plane.

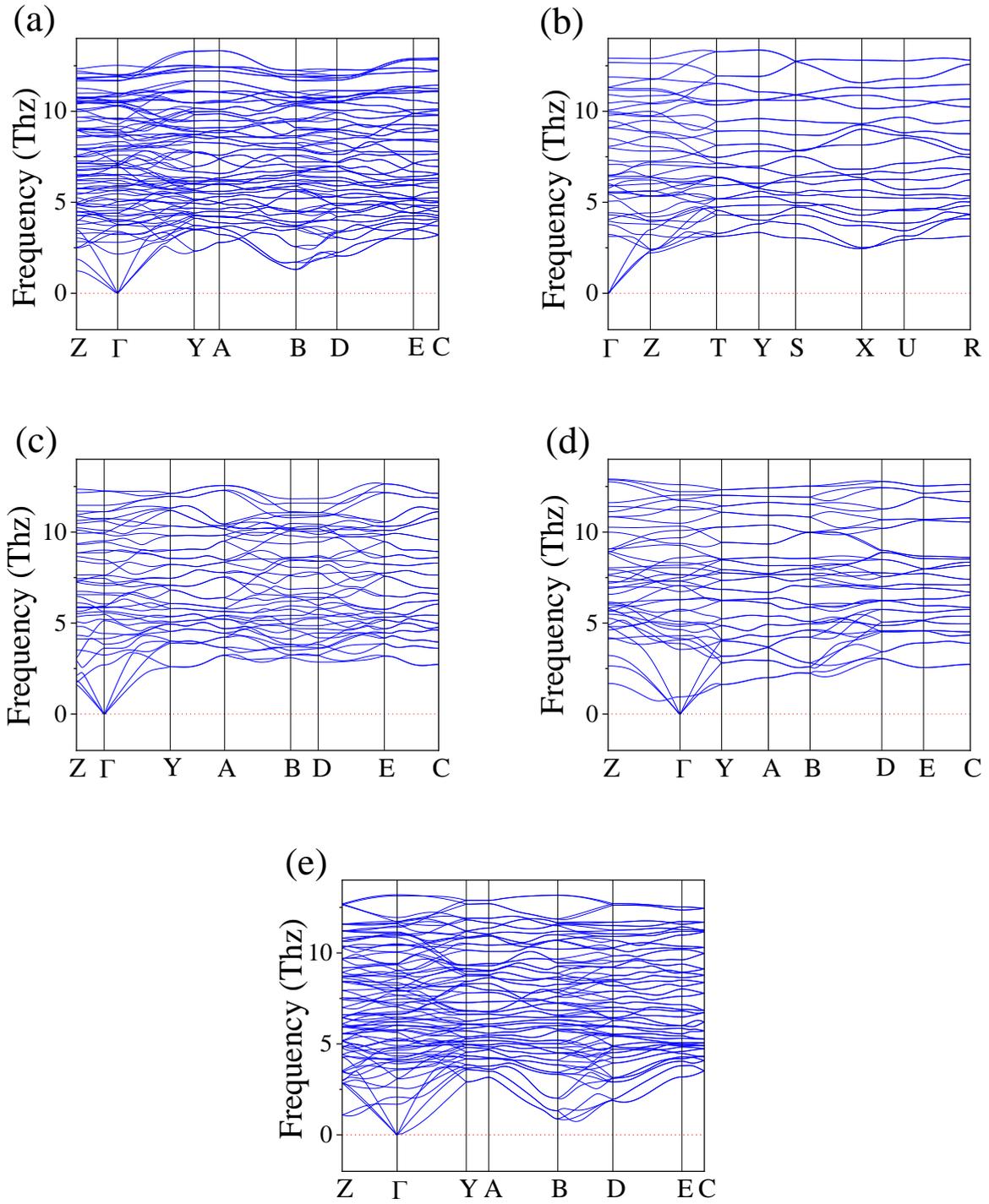

**Fig. 3.** Phonon spectra of the (a) *C*2/*m*-1, (b) *Pmmn*, (c) *P*2$_1$/*m*, (d) *C*2/*m*-2, and (e) *C*2/*c* phases of Ru$_2$Al$_5$.

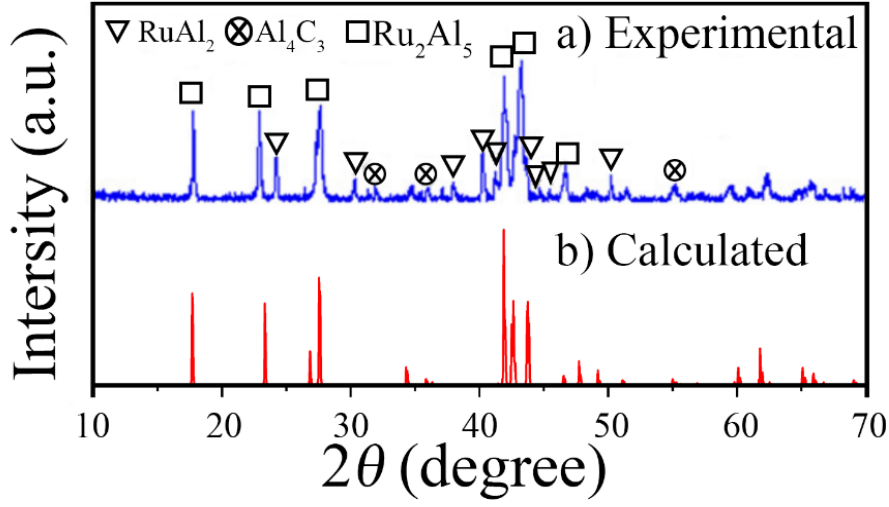

**Fig. 4.** (a) XRD patterns of milled and heat treated Al$_{79}$:Ru$_{21}$ powders and (b) simulated XRD pattern of the *Pmmn* structure. The experiment and calculation adopted the same wavelength of the light source (1.54 Å).

**Table 2.**

Calculated elastic constants $C_{ij}$ (GPa) of the five Ru$_2$Al$_5$ phases.

| Phases | $C_{11}$ | $C_{22}$ | $C_{33}$ | $C_{44}$ | $C_{55}$ | $C_{66}$ | $C_{12}$ | $C_{13}$ | $C_{15}$ | $C_{23}$ | $C_{25}$ | $C_{35}$ | $C_{46}$ |
|---|---|---|---|---|---|---|---|---|---|---|---|---|---|
| *C2/m*-1 | 251 | 275 | 274 | 76 | 49 | 107 | 127 | 60 | −6 | 6 | 32 | −31 | −2 |
| *Pmmn* | 266 | 234 | 234 | 90 | 80 | 87 | 104 | 56 | | 90 | | | |
| *P2$_1$/m* | 248 | 288 | 225 | 66 | 92 | 105 | 104 | 61 | -19 | 77 | 5 | −2 | 8 |
| *C2/m*-2 | 247 | 230 | 240 | 68 | 66 | 91 | 75 | 62 | 28 | 112 | 8 | 26 | 19 |
| *C2/c* | 208 | 298 | 225 | 74 | 76 | 23 | 87 | 125 | −23 | 65 | 7 | −1 | −25 |

The mechanical properties of a polycrystalline material under different deformation conditions largely depend on the elastic moduli of the material, represented by the bulk modulus (*B*), Young's modulus (*E*), and shear modulus (*G*). Elucidating the elastic moduli is critical for designing and selecting materials to meet specific engineering needs. The bulk and shear moduli of the monoclinic and orthorhombic structures were calculated using the Voigt-Reuss-Hill approach [29]. The Young's modulus (*E*) and Poisson's ratio (*v*) are determined as follows:

$$B_V = \frac{1}{9}(C_{11} + C_{22} + C_{33}) + \frac{2}{9}(C_{12} + C_{13} + C_{23}), \quad (2)$$

$$G_V = \frac{1}{15}(C_{11} + C_{22} + C_{33} - C_{12} - C_{13} - C_{23}) + \frac{1}{5}(C_{44} + C_{55} + C_{66}), \quad (3)$$

$$B_R = [(s_{11} + s_{22} + s_{33}) + 2(s_{12} + s_{13} + s_{23})]^{-1}, \quad (4)$$

$$G_R = 15[4(s_{11} + s_{22} + s_{33}) - 4(s_{12} + s_{13} + s_{23}) + 3(s_{44} + s_{55} + s_{66})]^{-1}, \quad (5)$$

$$B = \frac{1}{2}(B_V + B_R), \quad (6)$$

$$G = \frac{1}{2}(G_V + G_R), \tag{7}$$

$$E = \frac{9BG}{3B+G}, \tag{8}$$

$$v = \frac{3B-2G}{6B+2G}, \tag{9}$$

where the $s_{ij}$ are the elastic compliance constants. The subscripts $V$ and $R$ represent the Voigt and Reuss approximations, respectively. Table 3 lists the calculated elastic moduli and Poisson's ratios of the five Ru$_2$Al$_5$ structures. The $B/G$ value evaluates the brittleness or ductility of a material and can also indicate bonding type [35]. Studies have shown that a change in bonding type induces a ductility-to-brittleness transition in materials [36]. A material with a $B/G$ below 1.75 is generally considered as brittle, whereas a material with $B/G$ above 1.75 is typically ductile and can withstand large plastic deformation before breaking. The $P2_1/m$ and $Pmmn$ phases are brittle, as evidenced by their $B/G$ ratio below 1.75, indicating high hardness and breakage at low plastic deformation under an external force. The remaining three monoclinic structures have large $B/G$ ratios (>1.75), clarifying their ductile behaviors and high resistance to stress fracture. Hardness is closely related to the shear modulus and Poisson's ratio of a material. Materials with high hardness have high shear moduli, meaning that their dislocation movement is limited and occurs only under a high shear stress. Meanwhile, materials with a low Poisson's ratio exhibit small lateral contraction when stretched, this phenomenon is often related to directional bonds in the material. A low Poisson's ratio enhances the shear modulus by limiting the lateral expansion under shear. The large shear moduli (86 and 83 GPa, respectively) and small Poisson's ratios (0.240 and 0.248, respectively) of the $P2_1/m$ and $Pmmn$ phases (Table 3) suggest high hardness of these structures. According to Chen's hardness model, the Vickers hardness values [37] of materials is given by $H_V = (2k^2G)^{0.585} - 3$, where $k = G/B$. The $P2_1/m$ and $Pmmn$ phases possess large hardness values (12.6 and 11.6 GPa, respectively).

    Elastic anisotropy is an important property of crystals in practical applications. For example, the elastic anisotropy of piezoelectric fiber composites (piezocomposites) is exploited in various aerospace applications, including active vibration control systems, adaptive blades, and flexible morphing wings [38]. The direction-dependent Young's moduli of the candidate Ru$_2$Al$_5$ structures are plotted in Fig. 5. An ideally isotropic material yields a spherical Young's modulus diagram. If the Young's modulus pattern deviates from a sphere, the material is harder or softer in one or more directions. The Young's modulus in a certain direction can be determined by measuring the distance

from the origin of the coordinate system to a corresponding point on the surface. As demonstrated in Fig. 5, the five Ru$_2$Al$_5$ structures are elastically anisotropic, although the *Pmmn* structure of Ru$_2$Al$_5$ shows smaller elastic anisotropy than the other candidates.

**Table 3**

Calculated bulk moduli *B* (GPa), shear moduli *G* (GPa), Young's moduli *E* (GPa), *B*/*G* ratios, Poisson's ratios *v*, and hardness values (GPa) of Ru$_2$Al$_5$.

| Phases  | B   | G  | E   | B/G  | v     | H$_V$ |
|---------|-----|----|-----|------|-------|-------|
| *C*2/*m*-1 | 145 | 74 | 190 | 1.95 | 0.281 | 8.3   |
| *Pmmn*  | 137 | 83 | 207 | 1.65 | 0.248 | 11.6  |
| *P*2$_1$/*m* | 136 | 86 | 212 | 1.59 | 0.240 | 12.6  |
| *C*2/*m*-2 | 131 | 73 | 186 | 1.79 | 0.264 | 9.4   |
| *C*2/*c* | 142 | 51 | 136 | 2.81 | 0.341 | 3.6   |

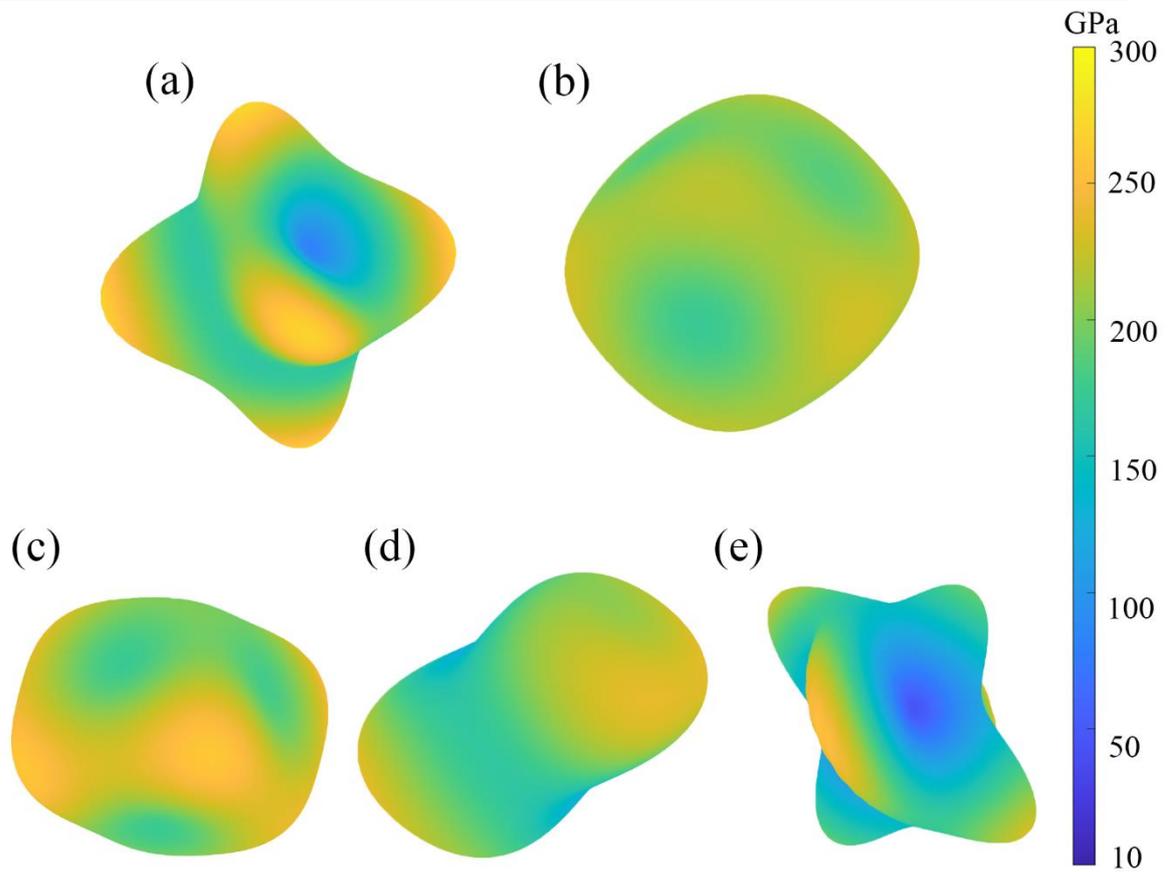

**Fig. 5.** Directional dependence of Young's moduli of the (a) *C*2/*m*-1, (b) *Pmmn*, (c) *P*2$_1$/*m*, (d) *C*2/*m*-2, and (e) *C*2/*c* phases of Ru$_2$Al$_5$.

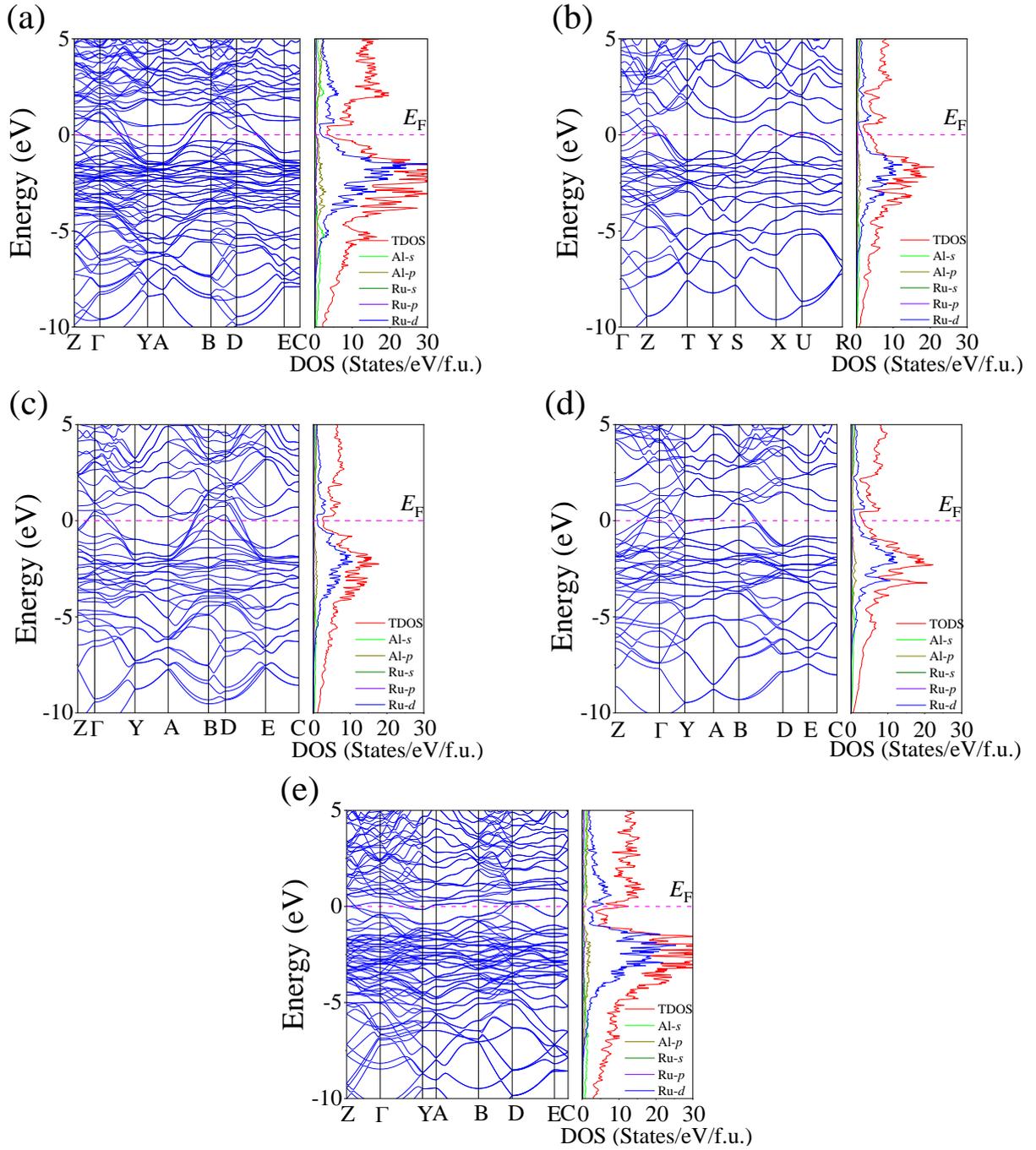

**Fig. 6.** Calculated band structures and the densities of stats of the (a) *C*2/*m*-1, (b) *Pmmn*, (c) *P*2$_1$/*m*, (d) *C*2/*m*-2, and (e) *C*2/*c* phases of Ru$_2$Al$_5$. Dashed lines represent the Fermi levels.

Fig. 6 plots the electronic energy bands and densities of states (DOSs) of the five Ru$_2$Al$_5$ structures. The valence band intersects the Fermi level with no band gap, indicating that electrons at the Fermi level can move freely through the conduction band. As the current flow is not restricted by a band gap, the five Ru$_2$Al$_5$ structures are electrically conductive, indicating that they are metallic compounds. Further examination of the atomic DOS shows that the Ru-*d* orbital electrons play a major role in the metallic properties of Ru$_2$Al$_5$. In addition, the total DOSs of all five structures feature a sharp valley around the Fermi level (known as a pseudogap), indicating the energy separation between the bonding and antibonding states. Earlier works [39-41] suggested that pseudogaps in DOSs indicate strong hybridization between the orbitals of the different atoms in a crystal structure, which contribute to structural stability. Clearly, the bonding states are nearly filled with the Fermi level around the pseudogap, underlying the structural stability of the $Ru_2Al_5$ compounds.

To determine the chemical bonding in the five Ru2Al5 compounds, we plot their charge-density distributions in Fig. 7. All structures exhibit regions of electron depletion and accumulation between the Al and Ru atoms, where ionic bonds are formed. Uneven chemical bond distribution affects the directionality, strength, bond lengths, and bond angles of the chemical bonds, thereby influencing the local stress state and symmetry of the crystal structure, and causing directional variations in the Young's modulus. The different elastic responses of the material in different directions ultimately manifest as an anisotropic Young's modulus. For example, the Young's modulus of *C*2/*m*-1 is strongly anisotropic along three main directions where the ionic bonds are concentrated. In the differential charge-density map of the *Pmmn* structure, the absolute values of charge density are considerably smaller along the (1 0 1), (1 0 −1), and (0 1 0) directions than along the (1 1 0), (1 −1 0), (0 1 1), and (0 1 −1) directions. Consequently, the three-dimensional Young's modulus is cubic with a 45° angle between the line connecting the centroid and the origin and the (0 0 1) plane.

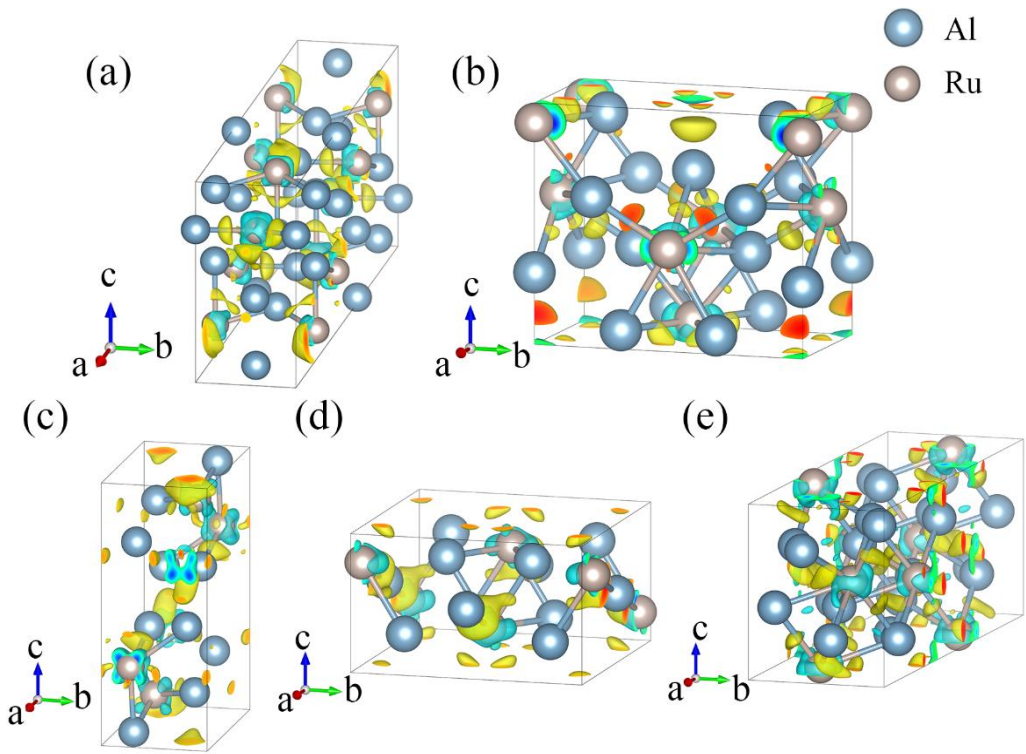

**Fig. 7.** Charge-density differences in the (a) *C*2/*m*-1, (b) *Pmmn*, (c) *P*2$_1$/*m*, (d) *C*2/*m*-2, and (e) *C*2/*c* phases of Ru$_2$Al$_5$. The blue and brown spheres represent Al and Ru atoms, respectively. Blue and yellow regions indicate electron depletion and accumulation, respectively.

## Conclusion

To solve the long-missing crystal structure of the experimentally synthesized Ru$_2$Al$_5$ intermetallic phase in Ru-Al alloys, we extensively searched the space of crystal structures using swarm intelligence structure simulations combined with first-principles calculations. Among the five stable phases of Ru$_2$Al$_5$ obtained by the search, the orthorhombic *Pmmn* structure yielded a calculated XRD spectra and lattice parameters that aligned well with the experimental data. Thus, the experimental phase of Ru$_2$Al$_5$ was confirmed as the *Pmmn* phase. The brittle/ductile behavior, elastic anisotropy, and ionic bonding of each Ru$_2$Al$_5$ candidate structure were demonstrated through elastic and electronic structural calculations.

**CRediT authorship contribution statement**

**Jing Luo:** Investigation, Data curation, Writing – original draft. **Meiguang Zhang:** Resources, Funding acquisition. Writing – review & editing. **Xiaofei Jia:** Investigation, Data curation. **Xuanmin Zhu:** Investigation. **Qun Wei:** Supervision, Project administration, Writing – review & editing.

**Declaration of competing interest**

The authors declare that they have no known competing financial interests or personal relationships that could have appeared to influence the work reported in this paper.

**Data availability**

Data will be made available on request.

## Acknowledgments

This work was financially supported by the National Natural Science Foundation of China (Grant Nos.: 11965005 and 11964026), the Natural Science Basic Research plan in Shaanxi Province of China (Grant Nos.: 2023-JC-YB-021, 2022JM-035), the Fundamental Research Funds for the Central Universities, and the 111 Project (B17035). All the authors thank the computing facilities at High Performance Computing Center of Xidian University.